\documentclass[letterpaper]{ptephy_v1}

\usepackage{subfig}
\usepackage{amsmath}
\usepackage{graphics}
\usepackage{url}

\usepackage{color}

\def\edth{{\rlap{$\partial$}\raise0.3em\hbox{$-$}}}
\newcommand{\bea}{\begin{eqnarray}}
\newcommand{\eea}{\end{eqnarray}}

\newcommand{\msun}{{M}_{\odot}}
\newcommand{\rsun}{{R}_{\odot}}
\newcommand{\zsun}{{Z}_{\odot}}

\begin{document}

\title{The possible existence of Pop III NS-BH binary and its detectability}


\author{\name{Tomoya Kinugawa}{1},
\name{Takashi Nakamura}{2}
and \name{Hiroyuki Nakano}{2}}

\address{${}^1$\affil{1}{Institute for Cosmic Ray Research, The University of Tokyo,
Chiba 277-8582, Japan}
\\
${}^2$\affil{2}{Department of Physics, Kyoto University, Kyoto 606-8502, Japan}
\\
}

\begin{abstract}
In the population synthesis simulations of Pop III stars,
many BH (Black Hole)-BH binaries with merger time less than
the age of the Universe $(\tau_{\rm H})$ are formed,
while NS (Neutron Star)-BH binaries are not.
The reason is that Pop III stars have no metal so that no mass loss is expected.
Then, in the final supernova explosion to NS, 
much mass is lost so that the semi major axis becomes too large
for Pop III NS-BH binaries to merge within $\tau_{\rm H}$.
However it is almost established that the kick velocity of the order
of $200-500{\rm~ km~s^{-1}}$ exists for NS from the observation of
the proper motion of the pulsar. Therefore, the semi major axis of the half
of NS-BH binaries can be smaller than that of the previous argument
for Pop III NS-BH binaries to decrease the merging time.
We perform population synthesis Monte Carlo simulations of Pop III NS-BH binaries
including the kick of NS and find that the event rate of Pop III NS-BH merger rate
is $\sim 1 {\rm Gpc^{-3} yr^{-1}}$.
This suggests that there is a good chance of the detection of Pop III NS-BH mergers
in O2 of Advanced LIGO  and Advanced Virgo from this autumn.

\end{abstract}
\subjectindex{E31, E32, E02, E01, E38}

\maketitle

\section{Introduction}

The proper-motion observations of pulsars show that
the pulsars had the kick velocity in the formation stage.
The young pulsars have proper velocity of
$200-500\rm~ km~s^{-1}$~\cite{Lyne:1994az,Hansen:1997zw}.
The physical mechanism of such kick velocity may be due to
the Harrison--Tademaru mechanism~\cite{Harrison:1975},
anisotropic emission of neutrinos, anisotropic explosion and so on
(see Lorimer~\cite{Lorimer:2008se} for the review).
Therefore, it is also reasonable to assume the existence of the proper motion
of the pulsars in the formation process of Pop III NSs,
although there is no direct evidence since no Pop III star or pulsar is observed.  
While, Repetto et al.~\cite{Repetto:2012bs} suggest that 
BHs also have a natal kick velocity comparable to pulsars from
the galactic latitude distribution of the low mass X-ray binaries in our galaxy.
But, first, this is not the direct observation of proper motion of BHs,
and second, since the mass of Pop III BHs is larger than Pop I and Pop II BHs,
their kick velocity might be so small that it can be neglected.
Therefore, we take into account the natal kick for Pop III NSs but not
for Pop III BHs in this paper.
The kick speed $v_{\rm k}$ obeys a Maxwellian distribution as
\begin{equation}
P(v_{\rm k})=\sqrt{\frac{2}{\pi}}\frac{v_{\rm k}^2}{\sigma_{\rm k}^2}\exp\left[-\frac{v_{\rm k}^2}{\sigma_{\rm k}^2}\right] \,,
\end{equation}
where $\sigma_{\rm k}$ is the dispersion.
The details of the method how to calculate the natal kick are shown
in Ref.~\cite{Hurley:2002rf}.

In this paper, we perform population synthesis Monte Carlo simulations
of Pop III binary stars.

\section{Brief Explanation of Our Population Synthesis Monte Carlo Simulations}

We calculate the Pop III NS-BH and Pop I and II NS-BH for comparison.
Pop I and Pop II stars mean solar metal stars and metal poor stars
whose metallicity is less than 10\% of solar metallicity, respectively. 
In this paper, we consider five metallicity cases of
$Z=0$ (Pop III), $Z=10^{-2}\zsun,~10^{-1.5}\zsun,~10^{-1}\zsun$
and $Z=10^{-0.5}\zsun,~Z=\zsun$ (Pop I).
There are important differences between Pop III and Pop I and II.
Pop III stars are (1) more massive, $>10~\msun$,
(2) smaller stellar radius compared with that of Pop I and II,
and (3) no stellar wind mass loss. 
These properties play key roles in binary interactions.

In order to estimate the event rate of NS-BH mergers and
the properties of NS-BH, we use the binary population synthesis
method~\cite{Hurley:2002rf,Kinugawa:2014zha,Kinugawa:2015nla}
which is the Monte Calro simulation of binary evolution.
First, we choose the binary initial conditions such as the primary mass $M_1$,
the mass ratio $M_2/M_1$, the separation $a$, 
and the eccentricity $e$ when the binary is born.
These binary initial conditions are chosen by the Monte Calro method
and the initial distribution functions such as the initial mass function (IMF), 
the initial mass ratio function (IMRF), the initial separation function (ISF),
and the initial eccentricity distribution function (IEF).
We adopt these distribution functions for Pop III stars and Pop I and II stars
as Table~\ref{IDF}.
\begin{table*}
\caption{The initial distribution functions in this paper.
}
\label{IDF}
\begin{center}
\begin{tabular}{ccc}
\hline
  & Pop III & Pop I,II\\
 \hline
 IMF& flat ($10~\msun<M<140~\msun$) & Salpeter ($5~\msun<M<140~\msun$)  \\
 IMRF &flat ($10/M<M_2/M_1<1$) &  flat ($0.1/M<M_2/M_1<1$)\\
ISF& logflat ($a_{\rm min}<a<10^6 \rsun$)&  logflat ($a_{\rm min}<a<10^6 \rsun$)\\
IEF& e ($0<e<1$)&e ($0<e<1$)\\
 \hline
\end{tabular}
\end{center}
\end{table*}
Second, we calculate the evolutions of the primary and secondary stars.
If the binary fulfills the condition of binary interaction,
we consider binary interactions such as the Roche lobe overflow (RLOF),
the common envelope (CE) phase, the tidal effect, the supernova effect,
and the gravitational radiation.
We treat these binary interactions as our previous studies
in Refs.~\cite{Kinugawa:2014zha,Kinugawa:2015nla}.
In this paper, we treat the binary interaction parameter such
as the CE parameter $\alpha\lambda$ and 
the lose fraction $\beta$ of transfered stellar matter during a RLOF
as $\alpha\lambda=1$, $\beta=0$, and 
the conservative core-merger criterion for all models~\cite{Kinugawa:2015nla}.
We adopt the maximum mass of NS is $3~\msun$
although this is near the maximum possible ones~\cite{Rhoades:1974fn,Hartle:1978}.
We calculate two kick velocity models of $\sigma_{\rm k}=265~\rm km/s$
and $\sigma_{\rm k}=500\rm~km/s$.

In order to calculate Pop III binary population synthesis,
we use the fitting formulae of Pop III stellar evolution and 
binary population synthesis code of Refs.~\cite{Kinugawa:2014zha, Kinugawa:2015nla}.
To be more accurate, however, we rewrite the lifetime of the He-burning phase as
\begin{eqnarray}
t_{\rm{He}}~[{\rm yr}]=\begin{cases}
                      214996+ 543838~\left(\frac{M}{10~\msun}\right)^{-1}+ 64028.1\left(\frac{M}{10~\msun}\right)^{-2}+  569484\left(\frac{M}{10~\msun}\right)^{-3}\\
 \;\;\;\;\;\;\;\;\;\;\;\; \text{($10 ~\msun \le M < 50 ~\msun$)} \,,\\
                     -108776 +3213670\left(\frac{M}{10~\msun}\right)^{-1} -5080480 \left(\frac{M}{10~\msun}\right)^{-2}\\
 \;\;\;\;\;\;\;\;\;\;\;\; \text{($20 ~{\rm{M}}_{\odot} \le M \le 100 ~{\rm{M}}_{\odot}$)} \,,\\
                      \end{cases}
\end{eqnarray}
and define the lifetime of the He-shell burning phase as
\begin{eqnarray}
t_{\rm{HeS}}~[{\rm yr}]=\begin{cases}
                      54343.2-145088~\left(\frac{M}{10~\msun}\right)^{-1}+ 165889\left(\frac{M}{10~\msun}\right)^{-2} -5377.16\left(\frac{M}{10~\msun}\right)\\
 \;\;\;\;\;\;\;\;\;\;\;\; \text{($10 ~\msun \le M < 50 ~\msun$)} \,,\\
                     -145220+34409.8 \left(\frac{M}{10~\msun}\right)-864.4\left(\frac{M}{10~\msun}\right)^{2}\\
 \;\;\;\;\;\;\;\;\;\;\;\; \text{($20 ~{\rm{M}}_{\odot} \le M \le 100 ~{\rm{M}}_{\odot}$)} \,.\\
                      \end{cases}
\end{eqnarray}
Thus, we redefine the ignition time of the C burning
in Ref.~\cite{Kinugawa:2014zha} as
$t^{\rm b}_{\rm C}=t_{\rm H}+t_{\rm He}+t_{\rm HeS}$.

On the other hand, in the Pop I and Pop II cases,
we use the fitting formulae of Ref.~\cite{Hurley:2000pk} and 
the formulae of binary interactions
in Refs.~\cite{Kinugawa:2014zha,Kinugawa:2015nla}.
We take the magnetic braking in the Pop I and Pop II cases
into account, while not for the Pop III case since no magnetic field is usually
expected for Pop III
star~\cite{Pudritz:1989,Kulsrud:1996km,Widrow:2002ud,Langer:2002tx,Doi:2011qf}. 
We use the formulae of angular momentum loss by magnetic braking
in Ref.~\cite{Hurley:2002rf}.
The stellar wind mass loss is effective in Pop I and II stars,
while no mass loss in Pop III stars.
The stellar wind mass loss makes binary not only light but also wide.
In this paper, we treat the stellar wind mass loss for Pop I and II stars
as follows.
In the case of massive main sequence whose luminosity is more than
$4000~ L_{\odot}$, the strong mass loss is observed.
We use the formula of Refs.~\cite{Nieuwenhuijzen1990,Hurley:2000pk},
\begin{equation}
\dot{M}_{\rm NJ} =9.6\times 10^{-15} \left (\frac{ Z}{ Z_{\odot}}\right) \left (\frac{R}{ R_{\odot}}\right) ^{0.81}\left (\frac{L}{ L_{\odot}}\right) ^{1.24}\left (\frac{M}{\msun}\right) ^{0.16}~ M_{\odot}\rm~yr^{-1} \,.
\end{equation}
For red giant stars, we use the formulation of Ref.~\cite{Kudritzki1978},
\begin{equation}
\dot{M}_{\rm R}=4\times10^{-13}\eta \left (\frac{L}{L_{\odot}}\right)  \left (\frac{R}{ R_{\odot}}\right) \left (\frac{M}{\msun}\right) ^{-1}~M_{\odot}\rm~yr^{-1} \,,
\end{equation}
with $\eta=0.5$ where $\eta$ is the parameter set
by the observations of horizontal branch stars in globular clusters~\cite{Iben1983}.
For the asymptotic giant branch stars,
we apply the formula of Ref.~\cite{Vassiliadis:1993zz},
\begin{equation}
\log \left (\frac{\dot{M}_{\rm VW}}{ M_{\odot}\rm~yr^{-1}}\right) =-11.4+0.0125\left[P_0-100~{\rm{max}}\left (\frac{M}{\msun}-2.5,~0\right) \right] \,,
\end{equation}
where $P_0$ is Mira pulsation period as
\begin{equation}
\log P_0={\rm min}\left (3.3,~-2.07-0.9\log \left (\frac{M}{\msun}\right) +1.94\log \left (\frac{R}{ R_{\odot}}\right) \right) \,.
\end{equation}
If the giant star fulfills the Humphreys-Davidson limit
($L>10^5~{\rm L_{\odot}}$
and $ (R/{\rm R_{\odot}})(L/{\rm L_{\odot}}) ^{1/2}>10^5$)~\cite{Humphreys:1994zz},
the radiation pressure becomes too high and the stellar surface becomes unstable
so that  the strong mass loss occurs.
Giant stars which are near the Humphreys-Davidson limit are called as
luminous blue variable (LBV) stars.
The stellar wind mass loss rates of LBV stars are typically from
$10^{-5}$ to $10^{-4}~\msun~\rm yr^{-1}$, or $10^{-3}~\msun~\rm yr^{-1}$
in the extreme case of $\eta$ Car~\cite{Smith:2014txa}.
In this paper, we adopt the additional mass loss rate,
\begin{equation}
\dot{M}_{\rm LBV}=1.5\times10^{-4}~ M_{\odot}\rm~yr^{-1}
\end{equation}
for the giants beyond the Humphreys-Davidson limit
so that $\dot{M}=\dot{M}+\dot{M}_{\rm LBV}$~\cite{Hurley:2002rf,Belczynski:2009xy}.
After a violent mass loss, the envelope of giant evaporates
and the star becomes a naked He star such as Wolf-Rayet star.
We adopt the mass loss rate for Wolf-Rayet like stars as
\begin{equation}
\dot{M}_{\rm WR} =10^{-13}L^{1.5} (1.0-\mu) \left (\frac{Z}{Z_{\odot}}\right) ^{0.86}~\rm M_{\odot}~yr^{-1} \,,
\end{equation}
where $\mu$ is 
\begin{equation}
\mu=\left (\frac{M-M_{\rm c}}{M}\right) {\rm min}\biggl\{5.0,~{\rm max}\biggl[1.2,~\left (\frac{L}{7.0\times10^4~ L_{\odot}}\right) ^{-0.5}\biggr]\biggr\}
\,.
\end{equation}
This Wolf-Rayet mass loss rate $\dot{M}_{\rm WR}$ is a combination
of the wind mass loss rate in Ref.~\cite{Hurley:2000pk} and
the metal dependent Wolf-Rayet wind in Ref.~\cite{Vink:2005zf}.  
For Wolf-Rayet stars, we use the mass loss formula $\dot{M}_{\rm WR} (\mu=0)$.

\section{Results}

In Table~\ref{NS-BH}, we show the number of NS-BH formations
and the number of NS-BHs which merge within 15 Gyrs for each metallicity
for the initial $10^6$ binaries.
The numbers are for the $\sigma_{\rm k}=265~\rm km/s$ models,
while the numbers in the parenthesis are
for the $\sigma_{\rm k}=500~\rm km/s$ models.
In the Pop I and Pop II cases,
the fraction of NS-BH formation and the fraction of merging NS-BH become larger
if the metallicity is lower.
In the Pop III case, the initial condition makes binaries easy
to be more massive compact objects. 
Thus, the fraction of NS-BH formation and the fraction of merging NS-BH
are larger than those for the Pop I and II cases. 
In all the metallicity cases, the numbers of NS-BH formation and
the numbers of merging NS-BH of the $\sigma_{\rm k}=265~\rm km/s$ model
are higher than those of the $\sigma_{\rm k}=500~\rm km/s$ model
due to disruption of the binary for the higher velocity kick.
In the case of Pop III in the no kick models,
almost all NS-BH cannot merge within the Hubble time
in our previous study~\cite{Kinugawa:2014zha,Kinugawa:2015nla},
because they eject a lot of mass at the supernova event,
and the separation becomes too wide due to the weak mass loss before the supernova.
Thus, in the Pop III case, they need supernova kick in order to merge
within the Hubble time.

\begin{figure}[!ht]
\begin{center}
 \includegraphics[width=0.45\textwidth,clip=true]{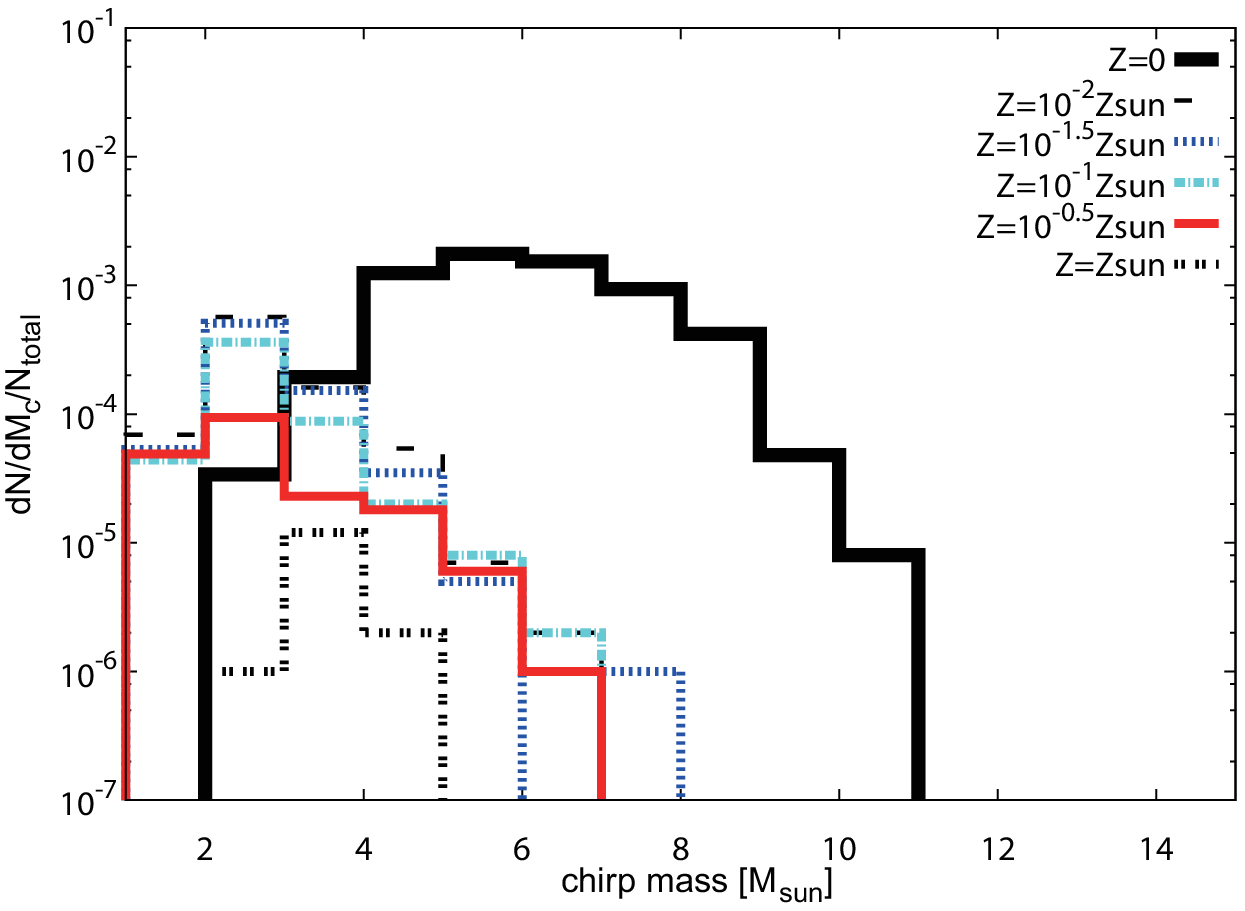}
 \includegraphics[width=0.45\textwidth,clip=true]{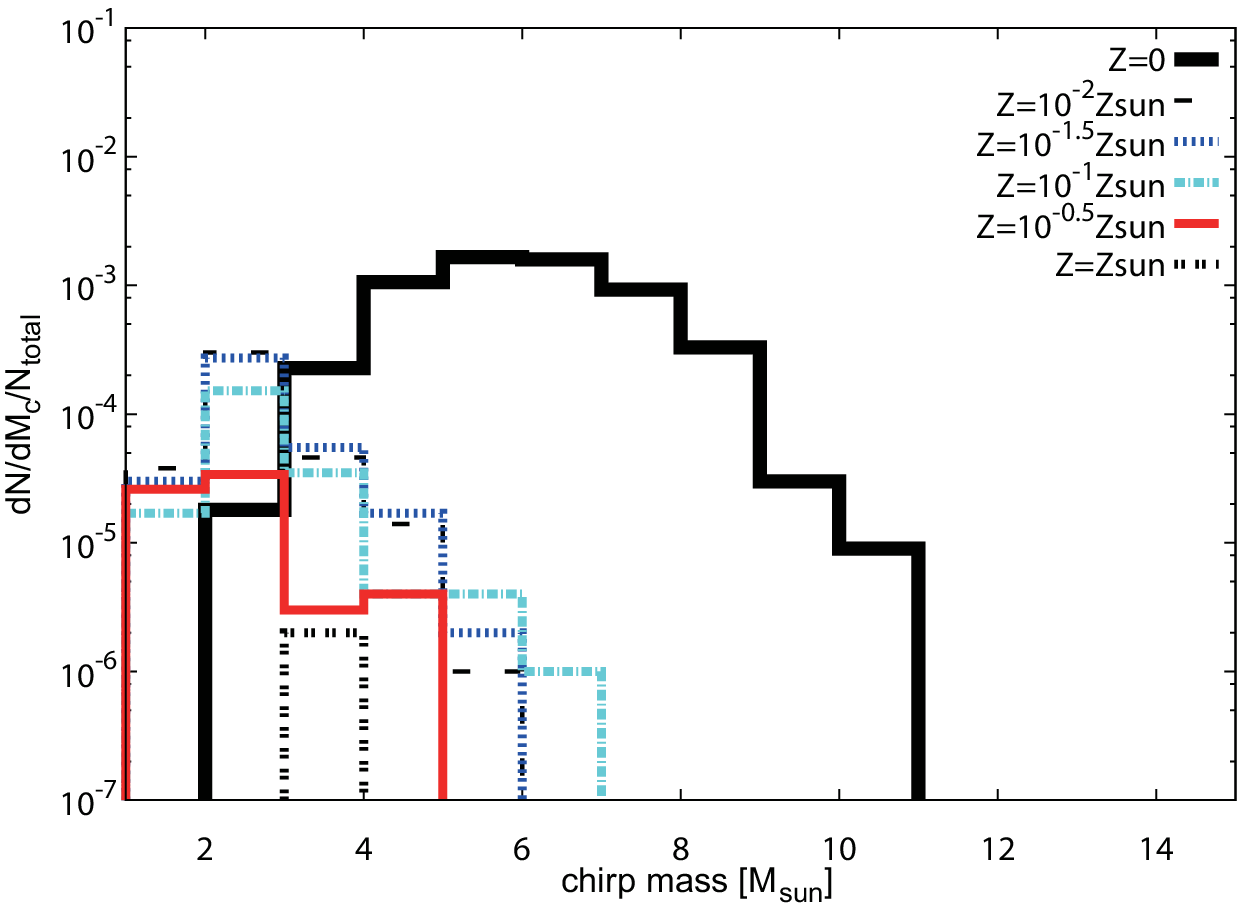}
\end{center}
 \caption{The chirp mass distributions of NS-BH for each metallicity.
The left and the right panels are the $\sigma_{\rm k}=265~\rm km/s$
and $\sigma_{\rm k}=500~\rm km/s$ models for each metallicity, respectively.}
 \label{fig:NS-BHchirpmass}
\end{figure}

Figure~\ref{fig:NS-BHchirpmass} shows the chirp mass distributions
of NS-BH which merge within 15 Gyr for each metallicity.
The left and right panels are the $\sigma_{\rm k}=265~\rm km/s$ model
and the $\sigma_{\rm k}=500~\rm km/s$ model for each metallicity, respectively. 
The chirp mass distributions for the Pop III case are clearly different from
those for the Pop I and II cases. 
The reasons are the difference of BH progenitor evolution
and the supernova mass ejection effect.
Pop III stars do not lose mass by the stellar wind and
the less binary interaction such as the common envelope phase.
Thus, the Pop III BHs tends to be more massive than those of Pop I and II.
In the case of Pop III NS progenitors, however, since they cannot lose
their mass before the supernova due to no wind mass loss and
the weak binary interaction,
they eject a lot of mass at the supernova event.
If the half of total binary mass is ejected at the supernova,
the binary disrupts for no kick velocity case.
Even though the binary does not disrupt, the orbit becomes wider
due to the mass ejection.
But, if the companion BH mass is massive, the effect is weak.
Thus, the chirp mass of Pop III NS-BH which merges within the Hubble time
is more massive than that of Pop I and II. 
The shapes of chirp mass distributions are changed a little
by the kick velocity values.
The peak values of chirp mass distributions, however, are not changed although the peak is not so sharp.

\begin{table*}
\caption{The number of NS-BH formations and the number of NS-BHs which merge
within 15 Gyrs for each metallicity for the initial $10^6$ binaries.
The numbers are for the $\sigma_{\rm k}=265~\rm km/s$ models,
while the numbers in the parenthesis are
for the $\sigma_{\rm k}=500~\rm km/s$ models.}
\label{NS-BH}
\begin{center}
\begin{tabular}{ccccccc}
\hline
 $Z$ & $\zsun$ & $10^{-0.5}\zsun$&$10^{-1}\zsun$&$10^{-1.5}\zsun$&$10^{-2}\zsun$&$0$\\
 \hline
 NS-BH& 148 (32) &598 (169) & 1296 (416)& 1686 (576)& 1896 (617)& 22638 (11192)\\
 merging NS-BH &15 (2)&191 (67)&525 (213)&755 (377)&862 (401)&9089 (5856)\\
 \hline
\end{tabular}
\end{center}
\end{table*}

We calculated the merger rate of NS-BH.
In order to calculate the merger rate, we need the star formation rate 
for each metallicity.
In the Pop III case, we use the star formation rate of Pop III
by Ref.~\cite{deSouza:2011ea}.
This SFR is calculated by a semi-analytical approach,
in which the following effects are considered: (1) the radiative feedback
on Pop III star formation, (2) the inhomogeneous re-ionization of
the intergalactic medium (IGM), and (3) the chemical pollution of the
IGM.~\footnote{
Recently, some researchers calculated the Pop III SFR
using the optical depth of Thomson scattering observed 
by Planck~\cite{Visbal:2015rpa, Hartwig:2016nde, Inayoshi:2016hco}.
Their SFRs are several or ten times lower than our SFR,
while they depend on the adopted optical depth,
the escape fraction of ionizing photons, IMF and the mass range.}
Using Pop III binary population synthesis results and the Pop III SFR,
the merger rate density of Pop III NS-BH 
$R_{\rm NS-BH}(t)~\rm [yr^{-1}~Gpc^{-3}]$ is calculated as
\begin{equation}
R_{\rm Pop III} (t) =\int^t_0\frac{f_{\rm b}}{ 1+f_{\rm b}}\frac{{ SFR} (t') }{<M>}\frac{N_{\rm NS-BH} (t-t') }{N_{\rm total}}dt' \,,
\end{equation}
where $f_{\rm b}$, $SFR(t')$, $N_{\rm NS-BH}(t-t')$, $<M>$ and $N_{\rm total}$
are the binary fraction, the star formation rate of Pop III at $t'$, 
the number of NS-BHs which merge from $t'$ to $t$,
the average mass, and the total number of the binary.
We use $f_b=0.5$, $<M>=75~\msun$, $N_{\rm total}=10^6$.
On the other hand, for the SFR of Pop I and II,
we use the SFR calculated by the observation~\cite{Madau:2014bja},
\begin{equation}
\Psi (z) =1.5\times10^{-2}\frac{ (1+z) ^{2.7}}{1+\left[\frac{1+z}{2.9}\right]^{5.6}}~\msun~\rm yr^{-1}~Mpc^{-3} \,, \label{popisfr}
\end{equation}
in $0<z \lesssim 8$. 
To decide the metallicity change as a function of the redshift,
we use the formula of galaxy mass-metallicity relation calculated by
simulation~\cite{Ma:2015ota},
\begin{equation}
\log\left (\frac{Z_*}{Z_{\odot}}\right) =0.40\left[\log\left (\frac{M_*}{M_{\odot}}\right) -10\right]+0.67\exp (-5.0z) -1.04 \,, \label{zmass}
\end{equation}
in $0<z<6$,
and the galaxy mass distribution fitted by the Shechter
function~\cite{Fontana:2006xg} as
\begin{equation}
\phi_{\rm sh} (M) dM=\phi^*\left (\frac{M}{M^*}\right) ^{\alpha}\exp\left (\frac{M}{M^*}\right) \frac{dM}{M^*} \,,\label{gmassfunc}
\end{equation}
where
\begin{align}
\phi^* (z) &=3.5\times10^{-3} (1+z) ^{-2.2} \,, \\
\log M^* (z) &=11.16+0.17z-0.07z^2 \,, \\
\alpha (z) &=-1.18-0.082z \,,
\end{align}
in $0<z<4$.
To calculate the metallicity evolution $z<8$,
we extrapolate Eqs.~\eqref{zmass} and \eqref{gmassfunc}. 
We decided the metallicity switching redshift as the intermediate value
in log as $Z=10^{-1.75}~(z=6.745),~Z=10^{-1.25}~(z=5.168),~Z=10^{-0.75}~(z=2.528)$, 
and $Z=10^{-0.25}~(z=0.096)$.
Using binary population synthesis results,
the SFRs and the metallicity,
the merger rates of each metallicity $R_{Z,\rm~ NS-BH}$ are given by
\begin{equation}
R_{Z,\rm~ NS-BH} (t) =\int^t_0\frac{f_{\rm b}}{1+f_{\rm b}}\frac{SFR (Z,t') }{<M>}\frac{\int^{140\msun}_{5\msun}IMF(M)dM}{\int^{140\msun}_{0.1\msun}IMF(M)dM}\frac{N_{Z,\rm~NS-BH} (t-t') }{N_{\rm total}}dt' \,.
\end{equation} 
where $f_b=0.5$ and $<M>=0.355~\msun$.

\begin{figure}[!ht]
\begin{center}
 \includegraphics[width=0.45\textwidth,clip=true]{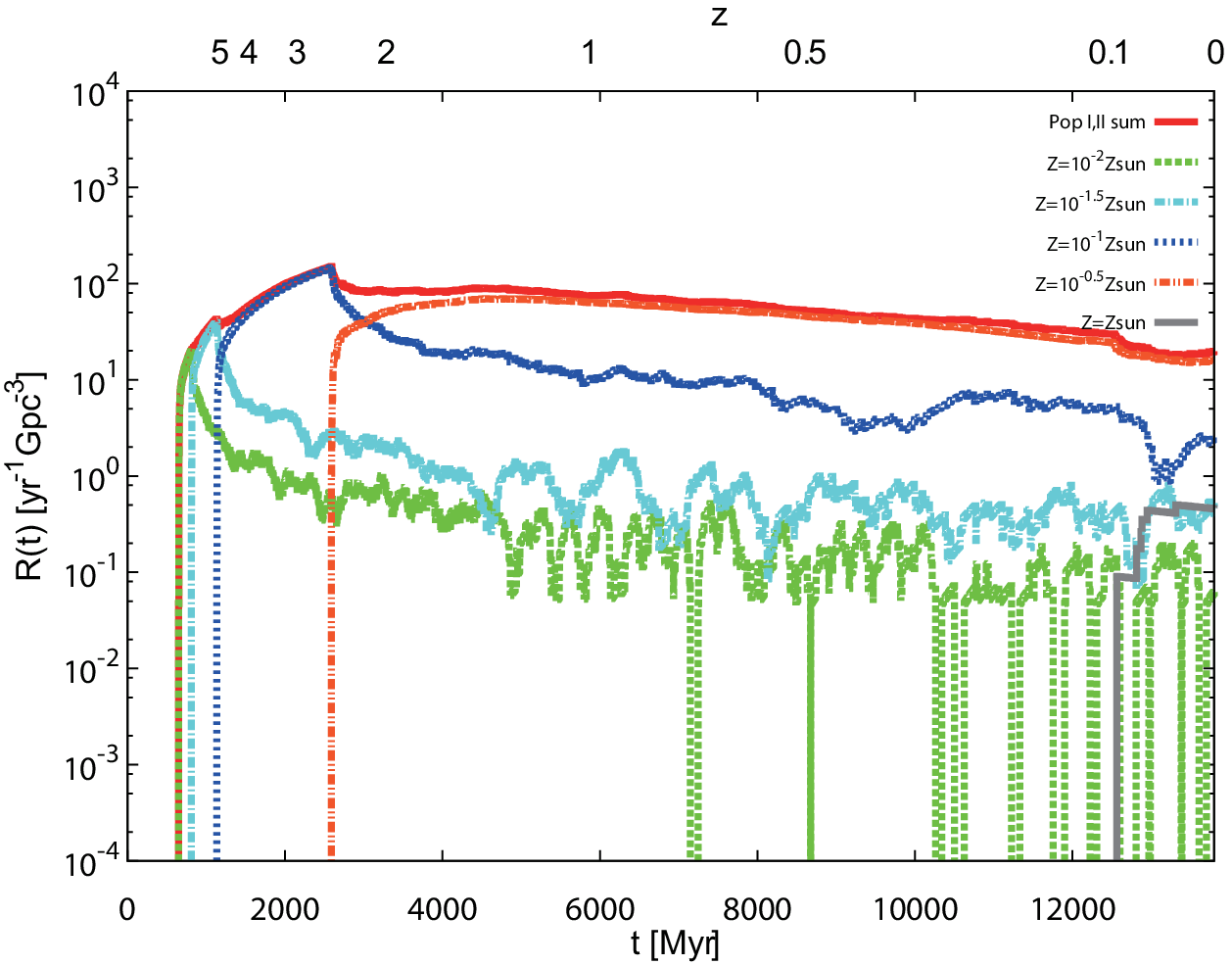}
 \includegraphics[width=0.45\textwidth,clip=true]{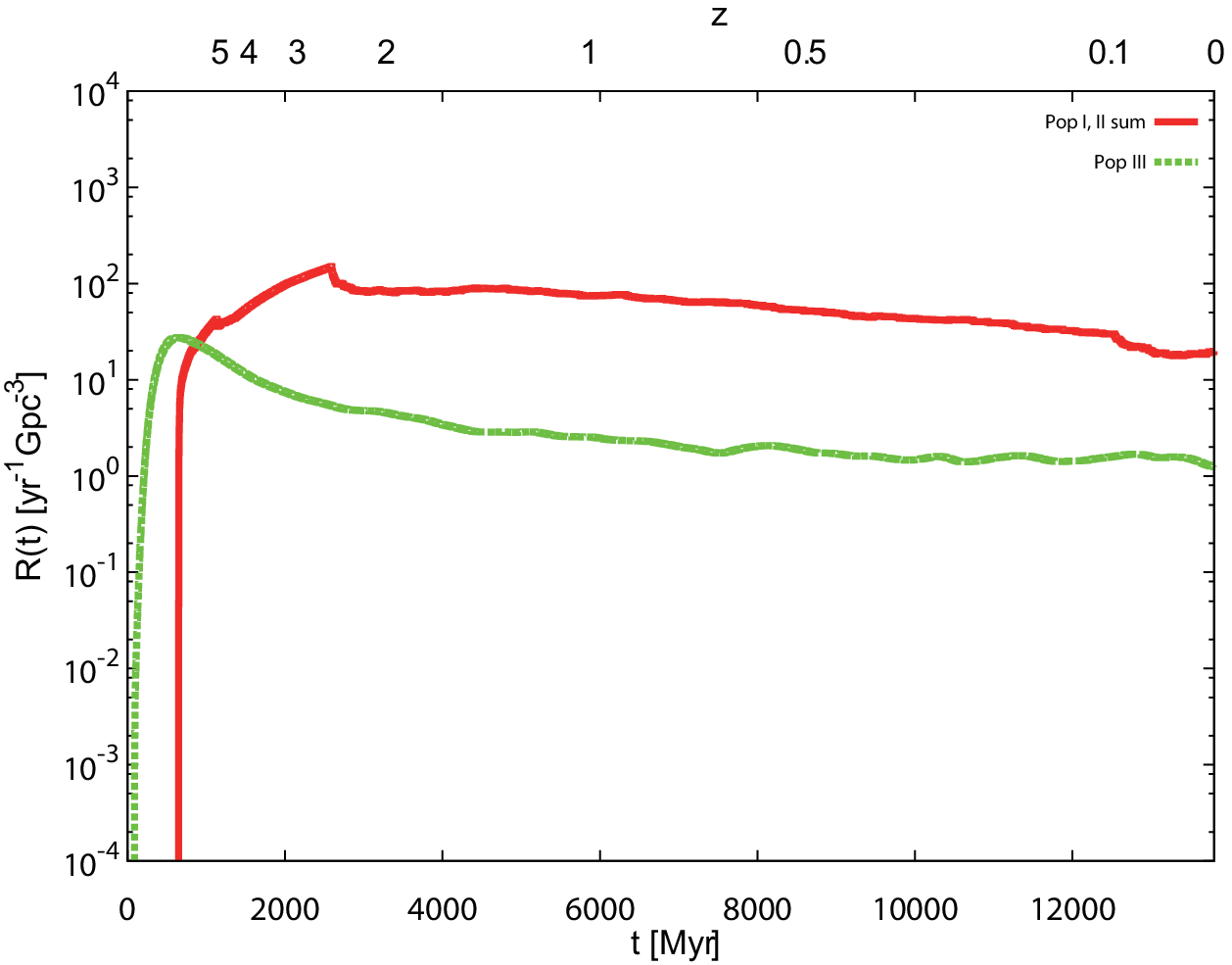}
\end{center}
 \caption{The merger rate density of NS-BH for the $\sigma_{\rm k}=265\rm~km/s$ model.
 The left panel shows the merger rate densities of Pop I and II,
 and Pop I and II summation. The right panel shows
 the merger rate densities of summation of Pop I and II, and Pop III.}
 \label{fig:NS-BHratelk}
\end{figure}

\begin{figure}[!ht]
\begin{center}
 \includegraphics[width=0.45\textwidth,clip=true]{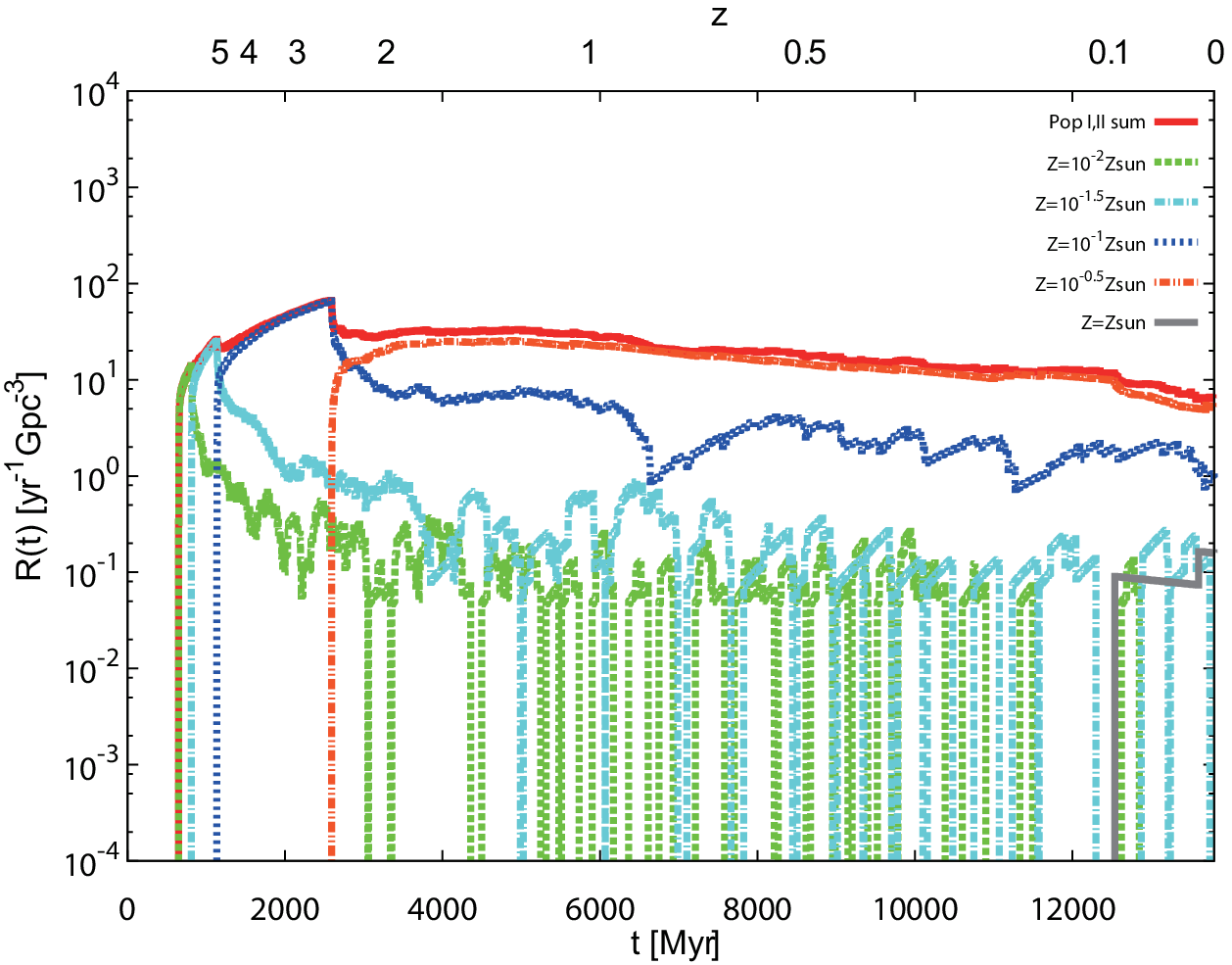}
 \includegraphics[width=0.45\textwidth,clip=true]{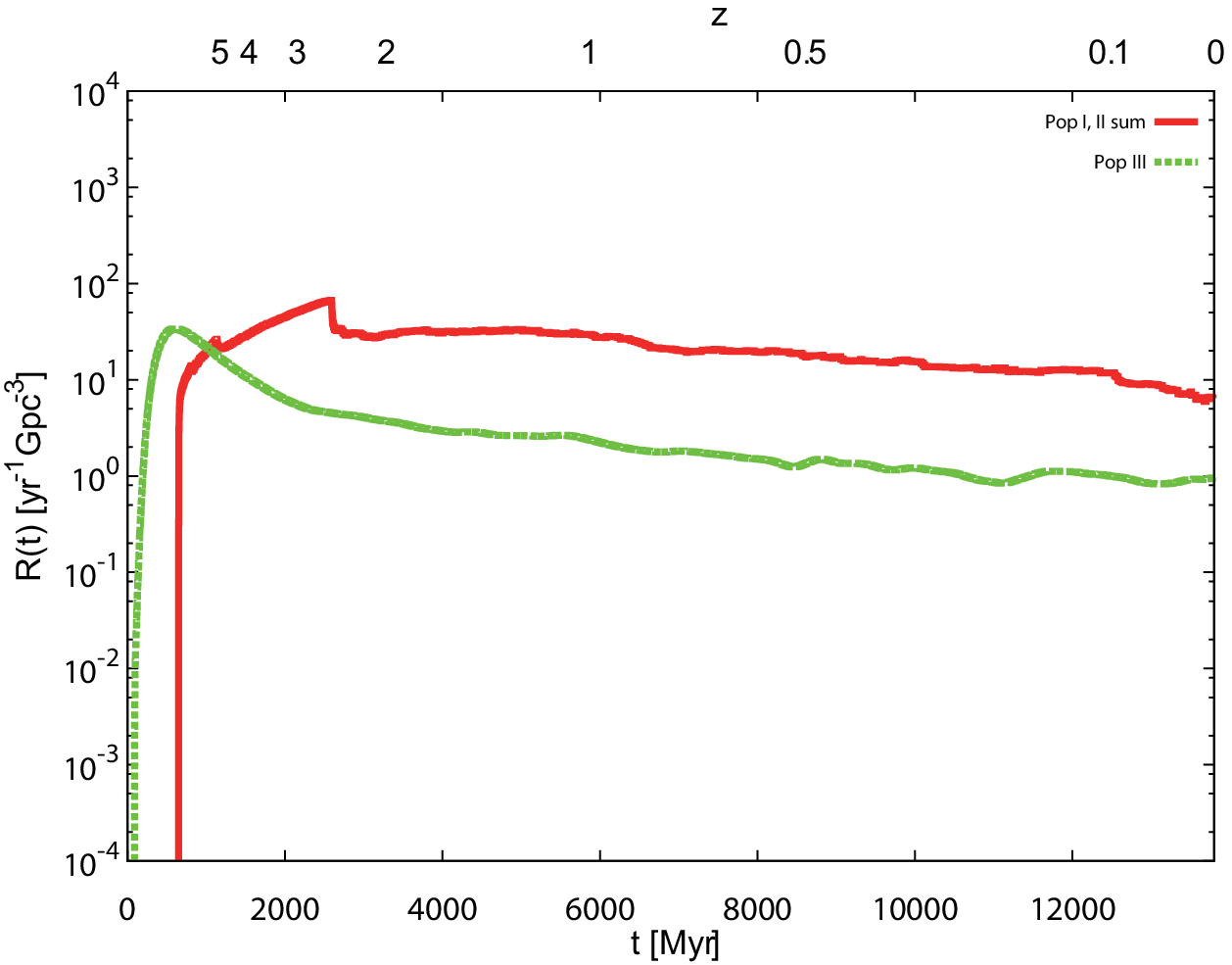}
\end{center}
 \caption{This figure is same as Fig.~\ref{fig:NS-BHratelk}
 but for the $\sigma_{\rm k}=500\rm~km/s$ model.}
 \label{fig:NS-BHratehk}
\end{figure}

Figures~\ref{fig:NS-BHratelk} and \ref{fig:NS-BHratehk}
present the merger rate densities of NS-BH.
Table~\ref{NS-BHrate} shows the NS-BH merger rates at the present day
for each metallicity and Pop I and II summation.
In the Pop I and II cases, $Z=10^{-0.5}\zsun$ contributes
the most of the merger rate. 
\begin{table*}
\caption{NS-BH merger rates at the present day [$\rm yr^{-1}~Gpc^{-3}$].}
\label{NS-BHrate}
\begin{center}
\begin{tabular}{cccccccc}
\hline
 $Z$ & $\zsun$ & $10^{-0.5}\zsun$&$10^{-1}\zsun$&$10^{-1.5}\zsun$&$10^{-2}\zsun$&Pop I,II sum&$0$\\
 \hline
 $\sigma_{\rm k}=265~\rm km/s$& 0.457 & 16.1 &2.57 & 0.523& 0.0623&19.7&1.25\\
 $\sigma_{\rm k}=500~\rm km/s$ &0.158&5.16&1.06&0&0&6.38&0.956\\
 \hline
\end{tabular}
\end{center}
\end{table*}


\section{Discussions}

\begin{figure}[!ht]
\begin{center}
 \includegraphics[width=0.45\textwidth,clip=true]{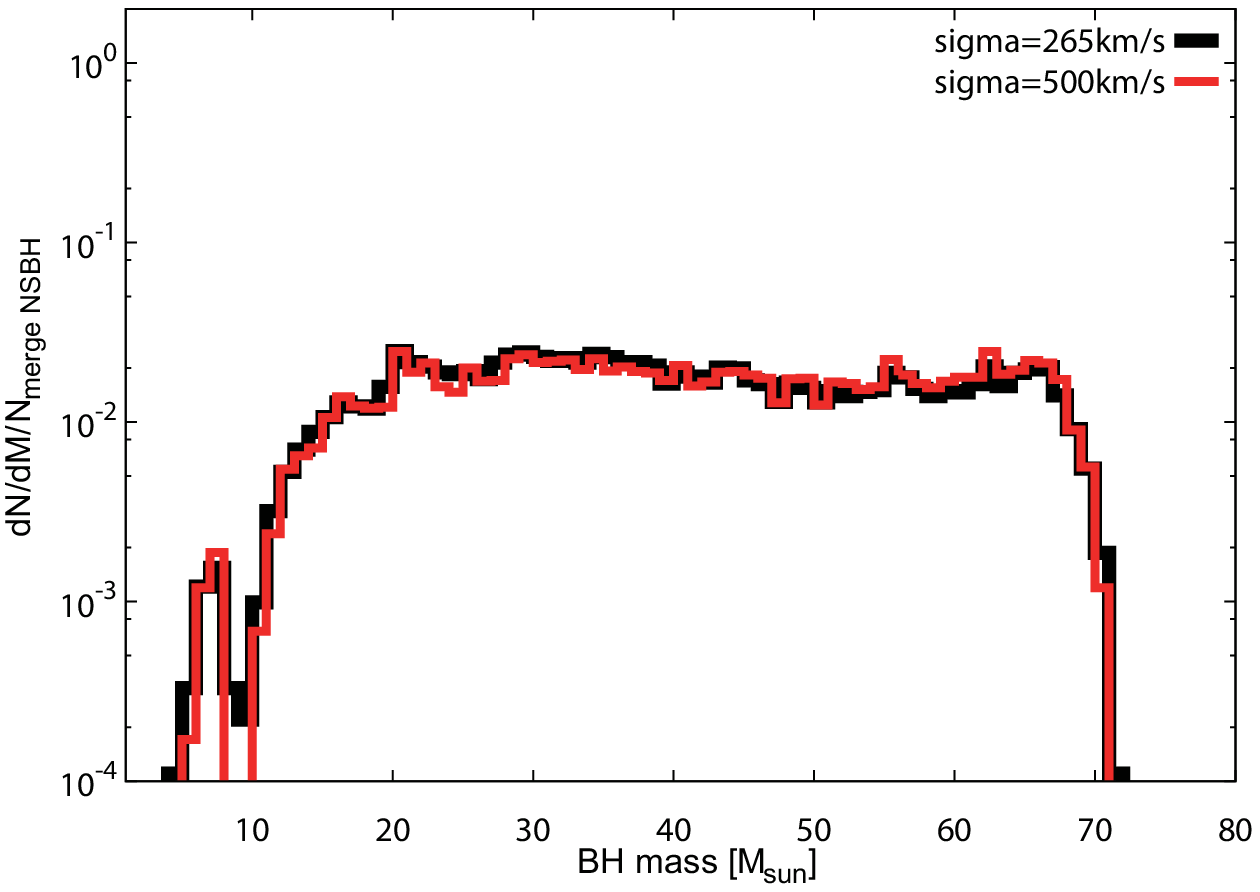}
 \includegraphics[width=0.45\textwidth,clip=true]{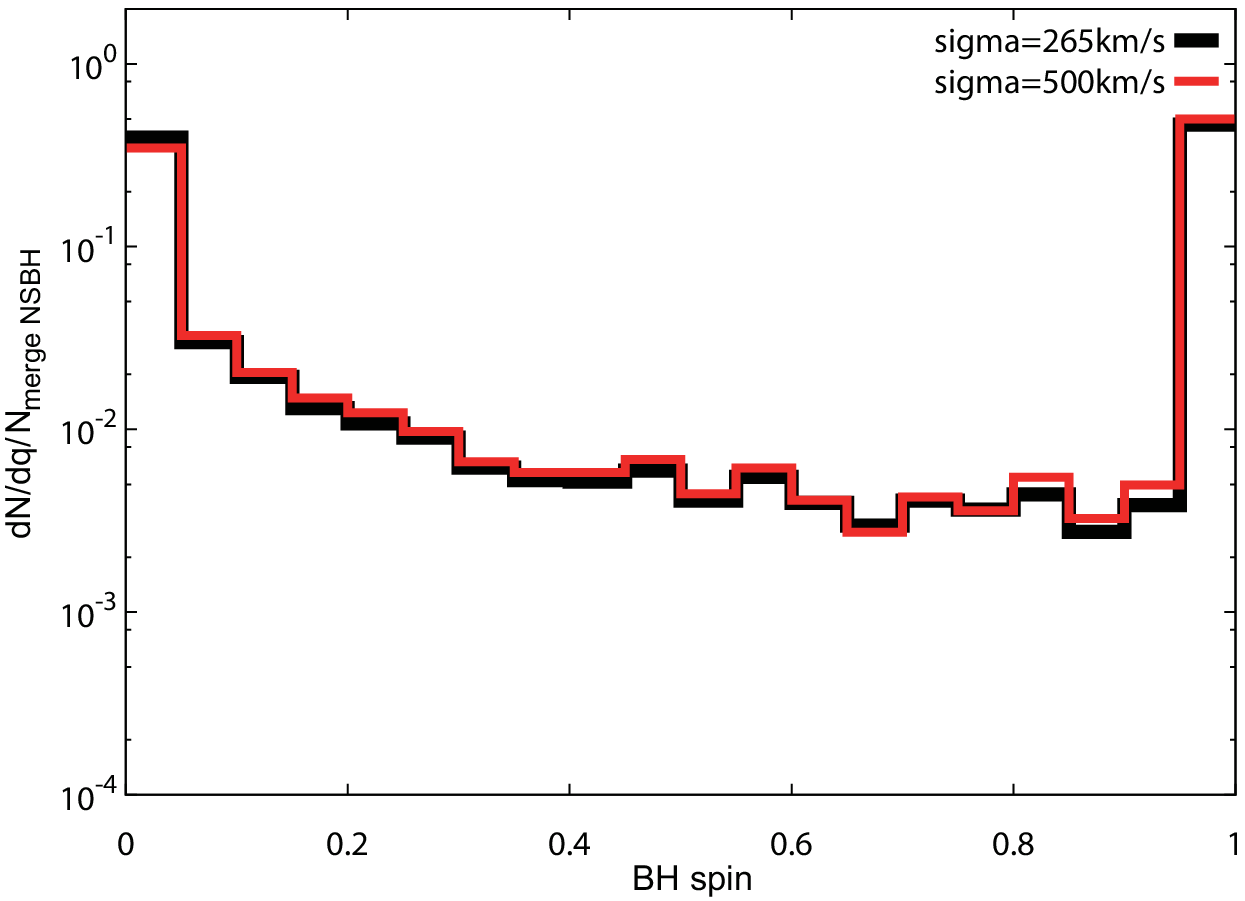}
\end{center}
 \caption{The left panel shows the BH mass of Pop III NS-BH which merge
 within 15 Gyr. The right panel shows the BH spin of Pop III NS-BH
 which merge within 15 Gyr.}
 \label{fig:BHmassspin}
\end{figure}

The NS-BH merger might be detected not only by a gravitational wave,
but also electromagnetic waves.
When a NS merges a BH, the NS is possibly disrupted by the tidal force from the BH.
At that time, the electromagnetic radiation is emitted due to the r-process.
This phenomenon is called as the kilonova.
In the case of massive BHs such as $\sim30~\msun$,
the kilonova does not occur generally since the NS is sucked into the BH
before the tidal disruption fully occurs.
However, if the spin of BH is high, it is conceivable
that the tidal disruption can occur.
Figure~\ref{fig:BHmassspin} shows the BH mass and spin distributions
of merging Pop III NS-BH.
The BH mass is massive, but a half of BHs have the extreme high spin.
Thus, they can produce the kilonova.
Of course, since the NS kick makes the orbit misalign with the spin of BH,
only the component of BH spin projected to the orbit
is effective~\cite{Kawaguchi:2015bwa}.
But, in the case of merging NS-BH, the kick velocity is low and
such misalignment effect is not so effective.
In order to consider the most misalign case, we assume that
there is no mass loss at the supernova, 
and that the direction of the natal kick is the direction orthogonal to the orbit. 
In order to avoid the disruption, the kick speed of the NS-BH progenitor
$v_{\rm k}$ should satisfy $v_{\rm esc}^2>v_{\rm o}^2+v_{\rm k}^2$
where $v_{\rm esc}$ and $v_{\rm o}$ are the escape speed and the orbit speed
before the supernova, respectively.
For simplicity, we assume no mass loss at the supernova.
The escape speed becomes $v_{\rm esc}^2=2v_{\rm o}^2$.
Thus, $v_{\rm k}<v_{\rm o}$ and the misalignment angle $\theta<45^\circ$.
If the spin of BH is $\sim1$, the effective spin is larger than $\cos45^\circ=0.7$.
Furthermore, actually the mass loss occurs at the supernova,
and $v_{\rm esc}$ is lower than that of no mass loss case.
Thus, the misalignment angle is much lower than that of no mass loss case
and the effective spin of BH is much larger than that of no mass loss case.
Therefore, the misalignment of kick is not effective. 

Finally, in Table~\ref{NS-BHevent}, we show the NS-BH event rates [$\rm yr^{-1}$]
by using the expected noise of Advanced LIGO.
Here, we used the fitting noise curve presented in Ref.~\cite{Keppel:2010qu}
for the design sensitivity which gives the average distance $\sim 200$ Mpc
for binary NS coalescences with two $1.4\msun$ NSs,
and for O2 where we assume the average distance $\sim 100$ Mpc
and half the design sensitivity~\cite{Aasi:2013wya}.
As for the NS-BH merger rates,
we adopted the values at the present day given in Table~\ref{NS-BHrate}
for $\sigma_{\rm k}=265~\rm km/s$.
The detailed calculation of the inspiral-merger-ringdown waveforms
is summarized in Ref.~\cite{Nakamura:2016hna},
based on Refs.~\cite{Dalal:2006qt,Ajith:2009bn}.
As the assumption of the chirp mass $M_c$ of the inspiral phase,
we fixed $M_c=2\msun$ (assuming $M_1=3.97\msun$ and $M_2=1.4\msun$)
and $6\msun$ (assuming $M_1=53.9\msun$ and $M_2=1.4\msun$)
for Pop I and II, and Pop III, respectively.
Although there is no contribution to the signal-to-noise ratio
(SNR) from the merger and ringdown phases for Pop I and II,
we find for Pop III that the ringdown phase contributes the SNR
and the difference in the remnant BH spin is also important.
Therefore, we give two cases, the BH spin $\sim 0$ and $1$
(see the right panel of Fig.~\ref{fig:BHmassspin}),
in Table~\ref{NS-BHevent}.
The detection rate of Pop III NSBH is larger than that of Pop I and II, although the Pop III NSBH rate depends  on the uncertaintyof the Pop III SFR.
The chirp mass of Pop III NSBH is cleary more massive than that of Pop I and II.
Thus, we might distinguish the Pop III NSBH from NSBH of Pop I and II.

\begin{table*}
\caption{NS-BH event rates for O2 by Advanced LIGO.
The numbers in the parenthesis denote those for the design sensitivity.
The maximum redshift $z_{\rm max}$
(and the luminosity distance $d_{L, {\rm max}}$) is obtained
by setting the averaged SNR $=8$.}
\label{NS-BHevent}
\begin{center}
\begin{tabular}{cccc}
\hline
 $Z$ & $z_{\rm max}$ & $d_{L, {\rm max}}$ [Mpc] & Event rate [$\rm yr^{-1}$] \\
 \hline
 Pop I, II sum & 0.0352 (0.0706) & 155 (318) & 0.269 (2.06) \\
 0, BH spin $\sim 0$ & 0.108 (0.216) & 500 (1070) & 0.445 (3.06) \\
 0, BH spin $\sim 1$ & 0.124 (0.264) & 580 (1340) & 0.658 (5.24) \\
 \hline
\end{tabular}
\end{center}
\end{table*}

\section*{Acknowledgment}

~~This work was supported by MEXT Grant-in-Aid for Scientific Research
on Innovative Areas,
``New Developments in Astrophysics Through Multi-Messenger Observations
of Gravitational Wave Sources'', No.~24103006 (TN, HN),
by the Grant-in-Aid from the Ministry of Education, Culture, Sports,
Science and Technology (MEXT) of Japan No.~15H02087 (TN),
and JSPS Grant-in-Aid for Scientific Research (C), No.~16K05347 (HN).


\end{document}